\documentstyle[psfig,floats,twocolumn,aps,prl]{revtex}
\def\pe{\langle PE \rangle}
\def\kvec{{\bf k}}
\def\kkvec{{\bf k'}}
\def\pvec{{\bf p}}
\def\ppvec{{\bf p'}}
\def\jvec{{\bf j}}
\def\rvec{{\bf r}}
\def\rrvec{{\bf r'}}
\def\w{\omega}
\def\Im{{\bf Im\,}}

\begin{document}
\ifx\undefined\psfig\def\psfig#1{ }\else\fi
\ifpreprintsty\else
\twocolumn[\hsize\textwidth%
\columnwidth\hsize\csname@twocolumnfalse\endcsname
\fi
\title{Dynamical Correlations in a Half-Filled Landau Level}
\author{Sergio Conti}
\address{Max-Planck-Institute for Mathematics in the Sciences, 
D-04103 Leipzig, Germany}
\author{Tapash Chakraborty$^\star$\nocite{byline}}
\address{Max-Planck-Institute for Physics of Complex Systems, 
D-01187 Dresden, Germany}
\maketitle
\begin{abstract}
We formulate a self-consistent field theory for the Chern-Simons 
fermions to study the dynamical response function of the quantum 
Hall system at $\nu=\frac12$. Our scheme includes the effect of 
correlations beyond the random-phase approximation (RPA) employed 
to this date for this system. We report results on the density 
response function, dynamic structure factor, the static structure 
factor, the longitudinal conductivity and the interaction energy 
of the system. The longitudinal conductivity calculated in this 
scheme shows linear dependence on the wave vector, like the
experimentals results and the RPA, but the absolute values are 
higher than the experimental results.

\end{abstract}
\ifpreprintsty\clearpage\else\vskip1pc]\fi
\narrowtext
Despite rapid progress in the field of quantum Hall effect in 
recent years, proper understanding of the state at half-filled 
Landau level still remains a challanging problem. A modified 
Fermi-liquid theory of Chern-Simons (CS) fermions, put forward by 
Halperin, Lee and Read (HLR) \cite{hlr} explained some of the 
anomalies observed in surface acoustic wave experiments (SAW) 
around the filling factor $\nu=\frac12$ \cite{willett}. 
One very interesting result of this theory was that at $\nu=\frac12$ 
the average effective magnetic field acting on the fermions vanishes 
and one expects a Fermi surface for those fermions. This result 
within the mean-field approach was later verified in experiments 
\cite{willett}, where one finds indications, albeit indirect, for 
the existence of a Fermi surface. In going beyond the mean field 
theory one has to include interactions via the Chern-Simons field 
in order to describe the dynamic response functions, transport 
properties, etc. HLR studied the response functions within the 
random phase approximation (RPA) which takes care of the direct 
Coulomb interaction and the fluctuations in Chern-Simons field. In 
the work of HLR the response functions were analyzed only in the 
long-wavelength limit. The RPA scheme was found to explain the wave 
vector dependence of the longitudinal conductivity derived from the 
SAW experiments \cite{hlr}. The absolute value of the calculated 
conductivity was, however, lower than the experimental results by a 
factor of two. The apparent success of HLR approach has marked the 
beginning of intense activities in the field, so much so that often 
the embellishments tend to overtake the actual facts. 

In this letter, we report our studies of the density response 
function, the dynamic structure factor, the static structure factor, 
and the longitudinal conductivity for the quantum Hall system at the 
filling factor $\nu=\frac12$, where we include correlations beyond 
the RPA scheme for the Chern-Simons fermions. In doing that, we have 
developed for the first time a variation on the theme of the 
celebrated self-consistent field theory of Singwi, Tosi, Land 
and Sj\"olander (STLS) \cite{stls} in the quantum Hall regime and 
at a half-filled Landau level. The efficacy of STLS approach over the
RPA scheme in describing correctly the dynamical properties is 
well established \cite{felde}. Our results for longitudinal conductivity 
show linear wave vector dependence, as in experiments and also in 
the RPA scheme, but the absolute values are higher than the 
experimental results. Most of the RPA results for the static and 
dynamical functions are also reported here for the first time.

We begin by presenting a few essential steps of the HLR approach to 
establish our notation. The CS transformation for spinless fermions is 
defined by (henceforth we use units where $\hbar=e=1$)
\cite{hlr}
\begin{equation}\label{eqCS}
\tilde\Psi^\dagger({\bf r})=\Psi^\dagger_e({\bf r})
\exp\left[-2{\rm i}\int d{\bf r}' \arg
({\bf r}-{\bf r}')\rho({\bf r}')\right]
\end{equation}
where $\Psi^\dagger_e({\bf r})$ is the electron creation operator,
$\tilde\Psi^\dagger({\bf r})$ is the transformed fermion operator, and 
$\arg ({\bf r})$ is the angle that vector ${\bf r}$ forms with the $x$-axis. 
The kinetic part of the Hamiltonian, which alters due to the transformation, 
is then
\begin{equation}
{\cal H}_{\rm kin}=\frac1{2m_b}\int d{\bf r}\,\tilde\Psi^\dagger({\bf r})
\left[-{\rm i}\nabla + \delta {\bf A}({\bf r})\right]^2\tilde\Psi({\bf r})
\label{kinetic}
\end{equation}
where $m_b$ is the electron band mass and
\begin{equation}
\delta A_i({\bf r}) = \int d{\bf r}' \phi_i(\rvec-\rrvec)\rho({\bf r}')
\end{equation}
[$\phi_i(\rvec)=2\nabla_i\arg(\rvec)$] is the CS field. Expanding the
right hand side of Eq.~(\ref{kinetic}) and keeping only up to
second-order contribution, one gets 
\begin{eqnarray}
{\cal H}=&-&\frac1{2m_b}\int d{\bf r}\,\tilde\Psi^\dagger({\bf r})
\nabla^2\tilde\Psi({\bf r}) \\
& + & \sum_{{\bf k}\ne0}{\rm i}v_1(k) j^T_{\bf k}
\rho_{-\bf k} + {1\over2} [v_0(k)+v_2(k)]\rho_{\bf k}\rho_{-\bf k}
\nonumber
\end{eqnarray}
where $j^T_\kvec = \hat\kvec\times\jvec_\kvec$ is the
transverse component of the transformed 
current operator
\begin{equation}
{\bf j}({\bf r})=\tilde\Psi^\dagger({\bf
  r})\frac{{\rm i}\nabla}{m_b}\tilde\Psi({\bf r}),
\end{equation}
(note that HLR used the diamagnetic current ${\bf J}+\rho_0{\bf A}/m_b$,
with $\rho_0$ being the equilibrium density), where
$v_0=2\pi/k$ is the Coulomb potential, $v_1(k)=4\pi/k$, and
$v_2(k)=(4\pi)^2\rho_0/m_bk^2$ [{\rm i}$\epsilon_{jh} \hat k_h v_1(k)$ is the
Fourier transform of $\phi_j(\rvec)$]. This 
Hamiltonian describes a system with the same density as the original system
where there is no magnetic field but contains a potential $v_1(k)$ 
which couples the density fluctuations to the transverse currents. This 
observation is the key ingredient for the exploitation of schemes that are 
normally applied to the electron gas in a zero magnetic field.

We intend to compute the response function matrix $\chi$, which gives
the density and transverse current responses $\rho(\kvec,\w)$ and 
$j_T(\kvec,\w)$ to external perturbation scalar
and transverse vector potentials $V^{\mathrm{pert}}$ and
$A_T^{\mathrm{pert}}$ via
\begin{equation}
  \left(\rho(\kvec,\w) \atop j_T(\kvec,\w)\right) = \chi(\kvec,\w) \cdot
  \left(V^{\mathrm{pert}}(\kvec,\w) \atop A_T^{\mathrm{pert}}(\kvec,\w)\right)
\end{equation}
where the longitudinal current and vector potential have been
eliminated using the continuity equation and gauge invariance.
Following the original derivation of STLS, we start from 
the equation of motion for the one-body Wigner distribution
function
\begin{equation}
f^{(1)} (\rvec, \pvec;t) = \sum_\kvec e^{{\rm i}\kvec\rvec}
\langle{a^\dagger_{\pvec-\kvec/2}(t) a^{}_{\pvec+\kvec/2}(t) }\rangle,
\end{equation}
which determines the density $\rho(\rvec,t)=\sum_\pvec f^{(1)}
(\rvec, \pvec;t)$ and the current  $\jvec(\rvec,t)=\sum_\pvec \pvec f^{(1)}
(\rvec, \pvec;t)/m_b$. 
In the semiclassical limit the Heisenberg equation of motion for the
electron operators $a_\kvec$ and $a^\dagger_\kvec$ gives
\begin{eqnarray}
&& {\partial\over\partial t} f^{(1)} (\rvec, \pvec;t)=
 {\pvec\cdot\nabla_\rvec\over m_b} f^{(1)} (\rvec, \pvec;t) \nonumber \\
 &&+\!\int\! d\rrvec\! \sum_\ppvec
 \left[  {(\pvec\!-\!\pvec')_j\over m_b}
 (\nabla_i \phi_j)(\rvec\!-\!\rrvec)  \nabla_{\pvec,i}
 {\vbox to 1.2\baselineskip{} }
 \right.\nonumber\\
 &&\left.
 +  [\nabla_i (v_0+v_2)](\rvec-\rrvec)\nabla_{\pvec,i}
 {\vbox to 1.2\baselineskip{} }\right]
 f^{(2)}(\rvec,\pvec; \rrvec,\ppvec; t) \nonumber\\
&& + \nabla_{\pvec,i} f^{(1)} \nabla_i
 V^{\mathrm{pert}}(\rvec,t)
 + {\pvec_j\over m_b} \nabla_{\pvec,i} f^{(1)} \nabla_i
 A_j^{\mathrm{pert}}(\rvec,t)
\label{eqf1dot}
\end{eqnarray}
where $\nabla_{\pvec,i}=\partial/\partial\pvec_i$, and
\begin{eqnarray}
&&f^{(2)} (\rvec, \pvec; \rrvec,\ppvec; t) 
= \sum_{\kvec,\kkvec}
e^{{\rm i}\kvec\rvec} e^{{\rm i}\kkvec\rrvec}\hskip2.2cm \nonumber\\
&&\hskip.8cm\times  
\langle{a^\dagger_{\pvec-\kvec/2}(t) a^{}_{\pvec+\kvec/2}(t)
a^\dagger_{\ppvec-\kkvec/2}(t) a^{}_{\ppvec+\kkvec/2}(t) }\rangle
 \end{eqnarray}
is the two-body distribution function. 
The key step in the STLS approximation consists in the decoupling
\begin{equation}\label{eqdecoup}
 f^{(2)}(\rvec,\pvec;\rrvec,\ppvec;t) \simeq
f^{(1)}(\rvec,\pvec;t)
f^{(1)}(\rrvec,\ppvec;t)  g(\rvec-\rrvec)
\end{equation}
i.e., in the assumption that the correlations in the perturbed, time-dependent 
state are identical to those in the unperturbed, equilibrium state, and are 
therefore described by the static pair correlation function $g(r)$. Notice 
that setting $g(r)=1$ in Eq.~(\ref{eqdecoup}) one recovers the RPA where 
short-range correlations are neglected.

Equation (\ref{eqf1dot}) is equivalent to the response of noninteracting
electrons to the effective potentials 
\begin{eqnarray}
\nabla_i  V^{\mathrm{eff}}&=& \nabla_i
V^{\mathrm{pert}} +
\int d\rrvec  \rho(\rrvec)
g(\rvec\!-\!\rrvec)\nabla_i(v_0+v_2)(\rvec\!-\!\rrvec) \nonumber\\
&& \label{eqeffpot1}
+ \sum_{j=1}^2\jvec^T_j(\rrvec) g(\rvec-\rrvec) \nabla_i 
\phi_j(\rvec\!-\!\rrvec)
\end{eqnarray}
and
\begin{equation}\label{eqeffpot2}
\nabla_i A_j^{\mathrm{eff}}= \nabla_i A_j^{\mathrm{pert}} +
{\cal P}_T \int d\rrvec  \rho(\rrvec)  g(\rvec-\rrvec) \nabla_i
\phi_j(\rvec\!-\!\rrvec) 
\end{equation}
where the continuity equation has been used to eliminate the
longitudinal part of the current, and ${\cal P}_T$ indicates projection
onto the subspace of transverse vector potentials.
In matrix notation this implies $\chi = \chi^0 [1 + U \chi]$, or
equivalently 
\begin{equation}\label{eqchi}
\chi = \chi^0\left[1-U\chi^0\right]^{-1},
\end{equation}
where 
\begin{equation}
\chi^0 = \left(\begin{array}{cc}
            \chi^0_{\rho\rho} & 0 \\
            0 & \chi^0_{T} \end{array} \right)
\end{equation}
is the ideal-gas response function which is known analytically
\cite{SternNCT98}. The matrix of the effective potentials, from
Eqs.~(\ref{eqeffpot1}--\ref{eqeffpot2}) is
\begin{equation}
U = \left(\begin{array}{cc}
            w_0(k)+w_2(k) & {\rm i}w_1(k)\\
        -{\rm i}w_1(k) & 0 \end{array} \right),
\end{equation}
where $w_\alpha(k)=[1-G_\alpha(k)] v_\alpha(k)$
and the local field factors $G_\alpha(k)$ are given by
\begin{equation}\label{eqgi}
  G_\alpha(k)=\sum_\pvec [1-S(\pvec)]
  {[\kvec\cdot(\kvec-\pvec)]^{(a_\alpha+b_\alpha)/2}
\over k^{a_\alpha} \left|\kvec-\pvec\right|^{b_\alpha}}
\end{equation}
with $a_0=b_0=1$, $a_1=b_1=2$, $a_2=0$, and $b_2=2$, and $S(k)$ is the 
static structure factor, i.e., the Fourier transform of the pair correlation 
function $g(r)$. Using the rotational invariance of $S(k)$ it is easy to show 
that in the $k\to0$ limit $G_0$ is linear in $k$, $G_2$ is quadratic in $k$, 
and $G_1$ has a finite limit, $G_1(0)=[1-g(0)]/2=1/2$. Notice that the RPA 
approximation of HLR amounts to $G_\alpha(k)=0$, i.e., 
$w_\alpha(k)=v_\alpha(k)$. 

The static structure factor entering Eq.~(\ref{eqgi}) is obtained from
the fluctuation-dissipation theorem,
 \begin{equation}\label{eqsqchi}
  S(k)=-{1\over\rho_0\pi}\int_0^\infty\Im\chi_{\rho\rho}(k,\w) d\w
 \end{equation}
where $\chi_{\rho\rho}(k,\w)$ is the density-density response
function given by Eq.~(\ref{eqchi}), 
\begin{equation}\label{eqchirhorho}
\chi_{\rho\rho}(k,\omega) = {\chi_{\rho\rho}^0(k,\omega)\over
 1 - \chi^0_{\rho\rho} \left[w_0(k) + w_2(k)
  + w_1(k)^2 \chi^0_T \right]}.
\end{equation}

Equations (\ref{eqchi}-\ref{eqchirhorho}) are then solved self-consistently 
for a given value of the dimensionless coupling strength, 
$r_s=r_0/a_B=2{e^2/\ell_0\omega_c}$, where $a_B=1/m_be^2$ is the
Bohr radius, $r_0=(\pi\rho_0)^{-1/2}$ is the average interparticle
spacing, $\ell_0=|B|^{-\frac12}$ is the magnetic length and $\omega_c=B/m_b$
is the cyclotron frequency. The relevant values of $r_s$ can be estimated in 
two ways: following 
HLR \cite{hlr} we can write, $r_s=2/C$, where $C\simeq0.3$ is related to the 
effective mass. Alternatively, we can obtain $r_s$ from a realistic estimate 
of $\omega_c$ and $e^2/\ell_0$, which is typically, $r_s=1\div 3$. The 
numerical results show little variation between the two cases but for 
definiteness we consider the first choice.

Since the density is not affected by the CS transformation of Eq.~(\ref{eqCS}), 
the density-density response function of the transformed system is identical 
to that of the original electron system, and therefore contains information 
about physical properties such as the structure factor, the conductivity, etc.
In the following we present and discuss our results for various
quantities derived from $\chi_{\rho\rho}$, and can therefore drop the
distinction between the two systems from now on. 

\begin{figure}[t]
  \begin{center}
    \leavevmode
    \psfig{figure=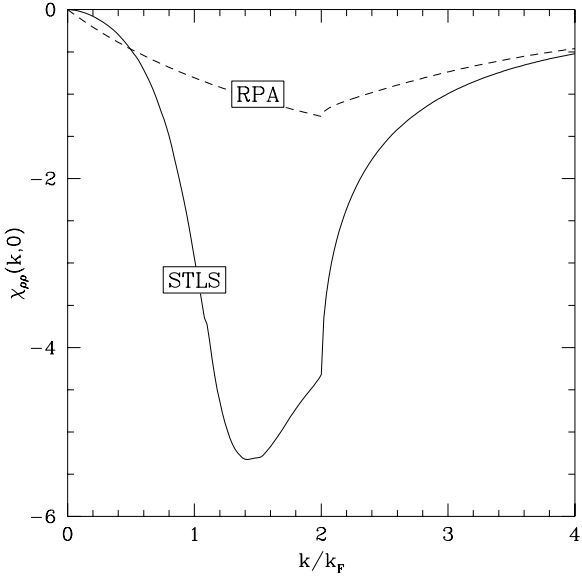,width=0.93\columnwidth}
  \end{center}
  \caption{Static density-density response function
    $\chi_{\rho\rho}(k,0)$ as a function of $k/k_F=k\ell_0$,
    calculated in the 
    RPA (dashed curve) and in STLS (full curve) for $r_s=7$. The
    discontinuity in 
    the derivative at $k=2k_F$ corresponds to the Fermi
    surface.}
  \label{fig:chi0}
  \label {fig1}
\end{figure}

In the static long-wavelength limit ($\omega\ll kv_F$, $k\ll k_F$,
$k_F=m_bv_F=1/\ell_0$ being the Fermi momentum) one has 
\begin{equation}\label{eqespstatchirhorho}
\chi^0_{\rho\rho}(k,\w)\simeq -{m_b\over 2\pi}\left(1+{\rm i}{\omega\over
 kv_F}\right)
\end{equation} and
\begin{equation}\label{eqespstatchiT}
\chi^0_T(k,\w)\simeq-{\rho_0\over m_b}\left(1+{\rm i}{2\omega\over kv_F}\right).
\end{equation} 
In the RPA, $v_2 = \rho_0 v_1^2/m_b$, hence the leading term in the
denominator of Eq.~(\ref{eqchirhorho}) cancels, and $\chi_{\rho\rho}(k,0)$ 
vanishes linearly for small $k$. If local field factors are included, this
cancellation does no longer take place and one gets
$\chi_{\rho\rho}(k,0)\simeq -(m_b/3\pi) k^2/k_F^2$. The results are presented 
in Figure \ref{fig1}, where one clearly sees the difference in the limit 
$k\to0$. We note that the $k^2$ dependence of the compressibility at 
$\nu=1/2$ has been observed also by other authors\cite{qsquare} in the dipole 
nature of $\nu=1/2$ state, which arises primarily due to projection to the 
lowest Landau level. Further, we find that excluding the cyclotron 
contribution, $\int \Im\chi_{\rho\rho} \w d\w\propto q^4$ and 
$\int \Im\chi_{\rho\rho} d\w\propto q^3$ apart from possible logarithmic terms.

The longitudinal conductivity, which is relevant to surface-acoustic-wave 
experiments is given by $\sigma_{xx}^{-1} = {\rm i} (k^2/\omega)
[\chi_{\rho\rho}^{-1}(k,\omega) - \chi_{\rho\rho}^{-1}(k,0)]$. Since
the speed of sound $c_s$ is small compared to the Fermi velocity
$v_F$ and $k\ll k_F$, we can use the limiting forms
(\ref{eqespstatchirhorho}-\ref{eqespstatchiT}) into
Eq.~(\ref{eqchirhorho}). This leads to the result
$\sigma_{xx}(k, c_sk) \simeq k/2\pi k_F$, which has the same linear
dependence on $k$, but is twice the experimental values\cite{willett}.
The RPA result of HLR is $\sigma^{RPA}_{xx}(k,c_sk) \simeq k/8\pi k_F$. A 
quantitative agreement with experiment can however be achieved if the 
CS interaction $\phi_j(\rvec)$ is softened at small separation. 

\begin{figure}[t]
  \begin{center}
    \leavevmode
    \psfig{figure=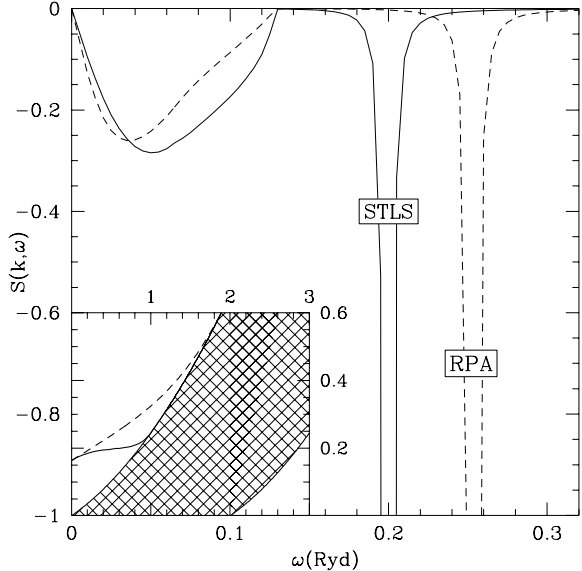,width=0.93\columnwidth}
  \end{center}
  \caption{Dynamic structure factor $S(k,\omega)$ for $k=0.6k_F$, as a
    function of $\omega$ (in Rydberg) in RPA (dashed curve) and STLS (full
    curve). The $\delta$-function peak corresponding to the inter-Landau-level 
    mode has been artificially broadened for clarity and contains most of the 
    spectral strength. The inset shows the excitation spectrum, composed of 
    the particle-hole continuum plus the sharp cyclotron mode.}
  \label{fig:skw}
  \label{fig2}
\end{figure}

Using the well-known asymptotic behaviors,
$\chi^0_{\rho\rho}=\rho_0k^2/m_b\omega^2$ and 
$\chi^0_T=O(k^2)$, valid for $k\ell_0\ll 1$, $kv_F \ll \omega$, one
sees that the dynamical structure factor 
\begin{equation}
  S(k,\omega)=-\frac1{\rho_0\pi}{\rm\bf Im}\,\chi_{\rho\rho}(k,\omega)
\end{equation}
has a pole at the cyclotron frequency $\omega=\omega_c$, describing
inter-Landau-level 
excitations, which -- at $k=0$ -- is unaffected by correlations, and
is in agreement with Kohn's theorem. The mode dispersion, which is computed
by locating the zeroes of the denominator in Eq.~(\ref{eqchirhorho}),
turns out to be significantly lower than the RPA result (see inset of
Fig. \ref{fig2}). 
Our finite-$k$ results are presented in Figure \ref{fig2} where we have a
$\delta$-function peak at $\omega\sim\omega_c$, corresponding to the
cyclotron motion and a continuum of particle-hole excitations in the range
\begin{equation}
\frac{k^2}{2m_b}-v_Fk \leq |\omega| \leq \frac{k^2}{2m_b}+v_Fk\,.
\end{equation}

\begin{figure}[t]
  \begin{center}
    \leavevmode
    \psfig{figure=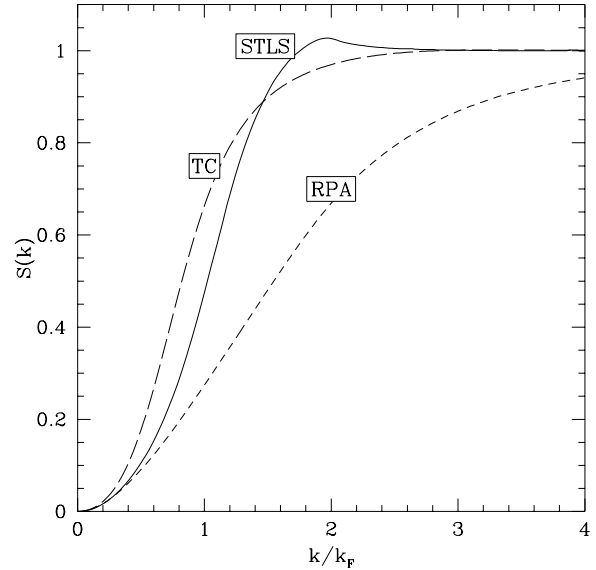,width=0.93\columnwidth}
  \end{center}
  \caption{Static structure factor $S(k)$ as a function of
    $k/k_F=k\ell_0$ in the RPA and the STLS scheme. The results are
    also compared with the results from Ref.~\protect\cite{tc} (TC).}
  \label{fig:sk1}
  \label{fig3}
\end{figure}

Our results for the static structure factor $S(k)$ are plotted in
Fig. \ref{fig3}. Here we compare our RPA results, calculated from
Eqs.~(\ref{eqsqchi}-\ref{eqchirhorho}) with $w_\alpha=v_\alpha$, and the 
STLS results, at $r_s=7$. These results are also compared with $S(k)$
calculated for a modified Laughlin state at $\nu=\frac12$ \cite{tc}, proposed 
by Read \cite{read}. All these curves obey the leading $(k\ell_0)^2/2$ 
behavior at small $k$. As expected, the STLS scheme includes substantial 
amount of correlations and hence is significantly higher than the RPA results 
near $k=2k_F$. 

Knowledge of the structure factor $S(k)$ allows us to compute the potential 
energy per particle, $\pe= {1\over2} \sum_k v_0(k) [S(k)-1]$. The full
interaction energy (defined as the total energy minus the noninteracting term
$\hbar\w_c/2$) is then obtained via coupling constant integration,
\begin{equation}
  E_{\rm int}(r_s) = {1\over r_s} \int_0^{r_s} dr_s' \langle PE(r_s') \rangle
\end{equation}
(we measure energies in units of $e^2/\ell_0$). The STLS result
$E_{\rm int}\simeq -0.48$ compares favorably with finite-size exact 
diagonalization studies\cite{Morf}, $E_{\rm int}=-0.466$. The RPA overestimates 
appreciably the interaction energy, and gives $E_{\rm int}\simeq -0.76$.  
At the same $r_s$, the STLS potential energy is $\pe=-0.49$, showing
that the inter-Landau-level kinetic energy is a minor contribution.

In summary, we have presented a self-consistent scheme for the
calculation of the dynamical response function of a quantum Hall fluid
at $\nu={1\over2}$, based on a generalization of the STLS method to
the case of Chern-Simons fermions. Our results exhibit significant
differences with the RPA computations, in particular on the longitudinal 
conductivity, the static response function and the structure factor. 

We wish to thank Peter Fulde for his kind hospitality at the
Max-Planck-Institute for Physics of Complex Systems in Dresden.


\begin{references}
\bibitem[\star]{byline}
On leave from: Institute of Mathematical Sciences, Taramani, Madras 
600 113, India
\bibitem{hlr} B.~I. Halperin, P.~A. Lee, and N. Read, Phys. Rev. B {\bf 47}, 
7312 (1993).
\bibitem{willett}
R.~L. Willett, Adv. Phys. {\bf 46}, 447 (1997).
\bibitem{stls}
K.~S. Singwi, M.~P. Tosi, R.~H. Land, and A. Sj\"olander, Phys. Rev. {\bf
179}, 589 (1968); K.~S. Singwi and M.~P. Tosi, Solid State Phys. {\bf 36},
177 (1981).
\bibitem{felde}
A. vom Felde, J. Spr\"osser-Prou, and J. Fink, Phys. Rev. B {\bf 40}, 10181
(1989). 
\bibitem{SternNCT98}
F. Stern, Phys. Rev. Lett. {\bf 18},  546  (1967);
R. Nifos\`\i, S. Conti, and M.~P. Tosi, cond-mat/9807085.
\bibitem{qsquare}
V. Pasquier and F.~D.~M. Haldane, cond-mat/9712169; R. Shankar and G. Murthy,
Phys. Rev. Lett. {\bf 79}, 4437 (1997); cond-mat/9802244; see also,
B.~I. Halperin and A. Stern, Phys. Rev. Lett. {\bf 80}, 5457 (1998).
\bibitem{tc}
T. Chakraborty, Phys. Rev. B {\bf 57}, 8812 (1998).
\bibitem{read}
N. Read, Semicond. Sci. Technol. {\bf 9}, 1859 (1994).
\bibitem{Morf}
R. Morf and N. d'Ambrumenil, Phys. Rev. Lett. {\bf 74}, 5116 (1995).

\end{references}
\end{document}